\documentclass{SCIS2024}

\usepackage{multirow}
\usepackage{setspace}
\usepackage[linesnumbered,ruled,vlined]{algorithm2e}
\begin{document}
\ArticleType{RESEARCH PAPER}
\Year{2024}
\Month{}
\Vol{}
\No{}
\DOI{}
\ArtNo{}
\ReceiveDate{}
\ReviseDate{}
\AcceptDate{}
\OnlineDate{}

\title{Coarse-to-Fine Lightweight Meta-Embedding for ID-Based Recommendation}{Coarse-to-Fine Sparse Meta-Embedding}

\author[1]{Yang WANG}{}
\author[1]{Haipeng LIU}{}
\author[1,2]{Zeqian YI}{}
\author[1]{Biao QIAN}{}
\author[1]{Meng WANG}{{eric.mengwang@gmail.com}}

\AuthorMark{Wang Y, et al.}

\AuthorCitation{Yang WANG, Haipeng LIU, Zeqian YI, Biao QIAN, Meng WANG}


\address[1]{School of Computer Science and Information Engineering, Hefei University of Technology, Hefei {\rm 230009}, China}
\address[2]{College of Information and Intelligence, Hunan Agricultural University, Changsha {\rm 410125}, China}
\abstract{The state-of-the-art recommendation systems have shifted the attention to efficient recommendation, \textit{e.g.,} on-device recommendation, under memory constraints. To this end, the existing methods either focused on the lightweight embeddings for both users and items, or involved on-device systems enjoying the compact embeddings to enhance reusability and reduces space complexity. However, they focus solely on the coarse granularity of embedding,  while overlook the fine-grained semantic nuances, to adversarially downgrade the efficacy of meta-embeddings in capturing the intricate  relationship over both user and item, consequently resulting into the suboptimal recommendations. In this paper, we aim to study how the meta-embedding can efficiently learn varied grained semantics, together with how the fine-grained meta-embedding can strengthen the representation of coarse-grained meta-embedding. To answer these questions, we develop a novel graph neural networks (GNNs) based recommender where each user and item serves as the node, linked directly to coarse-grained virtual nodes and indirectly to fine-grained virtual nodes, ensuring different grained semantic learning, while disclosing: 1) In contrast to coarse-grained semantics, fine-grained semantics are well captured through sparse meta-embeddings, which adaptively 2)  balance the embedding uniqueness and memory constraint. Additionally,  the initialization method come up upon SparsePCA, along with a soft thresholding activation function to render the sparseness of the meta-embeddings. We propose a weight bridging update strategy that focuses on matching each coarse-grained meta-embedding with several fine-grained meta-embeddings based on the users/items' semantics. Extensive experiments substantiate our method's superiority over existing baselines. Our code is available at \href{https://github.com/htyjers/C2F-MetaEmbed}{{https://github.com/htyjers/C2F-MetaEmbed}}.}


\maketitle

\section{Introduction}
Recommender systems (RSs)~\cite{2020Chen,2022Chen,2021Zhang} have revolutionized the way we navigate the overwhelming volume of available information and products by tailoring recommendations to individual user preferences. Benefiting from the collaborative filtering, RSs map users and items into high-dimensional latent spaces~\cite{2022Wang, 2021multi-modal, 2022Bayesian} for personalized recommendations via the strategies such as the dot product~\cite{FM}, multi-layer perceptrons~\cite{NCF}, and graph neural networks~\cite{Lightgcn}. However,with the data scale growing up,  complexity and resource demands of these systems heavily grows up. This issue is particularly highlighted in ID-based recommendation systems, where entities are used to uniformly represent all users and items, with each requiring a unique embedding vector. Consequently, traditional RSs suffer from substantial memory overhead, making them difficult to scale up. To address the challenge of optimizing memory usage without compromising recommendation accuracy, there has been a shift towards developing lightweight recommender systems~\cite{PEP,AutoEmb,ESAPN,OptEmbed,CIESS} via deep neural network~\cite{DHE,NimbleTT}.

One strategy is to construct the compact on-device session-based recommender systems~\cite{On-deviceRS, EODR, PPSR, NPOIR, 2024Long}, trained in the cloud and deployed on devices like smartphones. The primary challenge is to reduce the model size to fit resource-constrained devices. To tackle this, \cite{LEE} proposed to learn the elastic item embeddings, while adapting to device-specific memory constraints without retraining. Additionally, \cite{PEEL} proposed a personalized elastic embedding learning framework to on-device recommendations, by considering user and device heterogeneity. Moreover, \cite{ODNIR} suggested semi-tensor decomposition and self-supervised knowledge distillation~\cite{2023Qian, 2022Qian, 2024Unpacking} techniques for constructing ultra-compact on-device recommendation models. However, these on-device session-based recommenders face the daunting task of pinpointing the optimal embedding sizes and ensuring representational accuracy amidst an immense search space. Simultaneously, under memory constraints, learning from a teacher model necessitates denser embeddings, resulting in a higher likelihood of duplicate embeddings. Conversely, opting for a higher quantity of embeddings to reduce redundancy may lead to sparse embeddings, hindering sufficient knowledge to be captured from the teacher model.

To address this issue, parameter-sharing techniques have been proposed, such as hashing-based methods. The basic idea is to represent entities (such as user/item IDs) using a combination of meta-embedding vectors determined by hash functions, which can be shared across entities. By sharing parameters, hashing-based methods~\cite{QR, FBDH} significantly reduce the memory footprint while maintaining distinct representations for each entity. These solutions utilize one or several meta-embedding tables, also known as codebooks, comprising fixed-size embedding vectors. Each user/item ID is hashed into a combination of indexes, enabling the construction of entity embeddings by aggregating meta-embedding vectors drawn from these hashed indexes. This approach, known as compositional embedding, allows the meta-embedding tables to contain far fewer embedding vectors than a full embedding table, yet still offer the unique representations to each entity. \cite{CEL} proposed clustered embedding learning, where entities within the same cluster jointly learn and share an embedding, significantly reducing memory costs. \cite{LCCE} proposed representing each entity by combining a pair of sparse embeddings from two independent, substantially smaller meta-embedding tables, which are encouraged to encode mutually complementary information. \cite{LightweightEmbeddings} introduced a lightweight embedding framework dedicated to graph neural networks (GNNs) based recommenders, which can adaptively learn all meta-embeddings based on semantics and provide weighted meta-embedding assignments for each entity. However, existing meta-embedding methods tend to focus solely on coarse-grained learning, and fail to capture richer, fine-grained semantic nuances. This limitation significantly impacts the efficacy of meta-embeddings in representing the intricate information of the user and item, consequently resulting into the suboptimal recommendations. The above standing-out problem naturally promotes us to delve into the following questions \textit{how the meta-embedding can learn different grained semantics} and \textit{how the fine meta-embedding can strengthen the representation of coarse-grained meta-embedding} for ID-based recommendation.

As motivated above, we  propose a novel GNN-based model to facilitate multi-grained semantic learning for meta-embedding. Specifically, we construct a graph based on direct interactions between users and items. Following that, we design a two-tiered virtual node structure~\cite{Semi}. First, coarse meta-embedding nodes are superimposed, directly connecting to all real user and item nodes, thus capturing broad, shared semantics. Subsequently, fine-grained nodes are established, which exclusively link to these coarse-grained nodes, enabling the mapping of more detailed semantics. This hierarchical and selective connectivity ensures that the model learns wide-ranging semantic information at the coarse-grained level, while focusing on fine-grained personalized features. This coarse-to-fine learning strategy fosters a well understanding of the semantics of users/items from global and local side.

\begin{figure*}
  \centering
  \includegraphics[width=\textwidth]{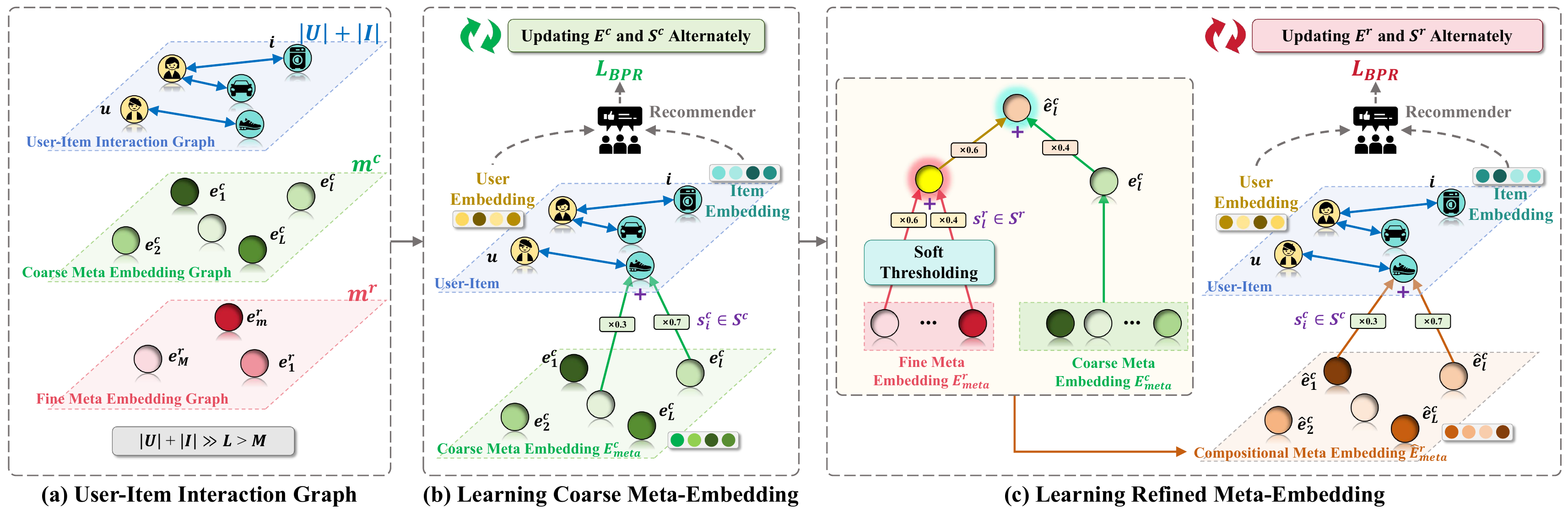}
  \caption{Illustration of the proposed pipeline. Our basic idea is to construct a the hierarchical and selective connectivity graph (a), which ensures that the model learns wide-ranging semantic information at the coarse level (b), while also focusing on granular, personalized features at the fine level (c)}\label{model}
\end{figure*}

To effectively combine the two phases of our embedding learning approach, we adopt a process of optimizing the coarse meta-embeddings as a springboard for developing sparse fine-grained embeddings. Through SparsePCA(Sparse Principal Component Analysis)~\cite{SPCA}, we selectively generate random entries within the coarse-grained meta-embedding codebook. This targeted initialization not only preserves the sparsity but also ensures the associative relationships between users and items. Moreover, we design the soft thresholding during training then dynamically adjusts sparsity levels. Unlike SparsePCA's rigid approach, soft thresholding adapts to data and enhancing the robustness. This combination establishes a finely-tuned, dynamic embedding process that remains sensitive to fine-grained semantics. Last but not least, a novel weight comes up to bridge update strategy that dynamically matches coarse-grained meta-embeddings with multiple fine-grained meta-embeddings based on semantic relevance. This innovation ensures that our method remains efficient and effective in representing complex information. Extensive experiments demonstrate that our approach outperforms the existing methods in both recommendation accuracy and memory efficiency, making it highly suitable for applications under memory constraints.The overall pipeline is shown in Fig.\ref{model}.

In summary, our contributions are summarized below:
\begin{itemize}
    \item[$\bullet$] We propose a novel GNN-based model for multi-grained semantics of meta-embeddings, the hierarchical and selective connectivity ensures that the model learns wide-ranging semantic information at the coarse level, while also focusing on granular, personalized features at the fine level.
    \item[$\bullet$] An initialization approach is adopted by optimizing the coarse meta-embeddings and SparsePCA to selectively seed the embedding codebook, which preserves the sparsity and maintains associative relationships between users and items.
    \item[$\bullet$] A soft thresholding technique is developed to dynamically adjust the sparsity levels during training, while offering a more adaptable and robust model that responds to the varied intricacies of the data.
    \item[$\bullet$] Lastly, we develop an innovative weight to bridge update strategy that dynamically aligns coarse-grained meta-embeddings with multiple fine-grained meta-embeddings based on semantic relevance, ensuring efficacy of the recommendation.
\end{itemize}

\section{Preliminaries}
Before shedding light on our technique, we elaborate the lightweight embedding for ID-based recommendation and the GNN-based meta-embedding learning.
\subsection{Efficient Meta-Embedding for ID-Based Recommendations}
In ID-based recommendation, with the user set $\mathcal{U}$ and the item set $\mathcal{I}$, the total number of entities is given by $N = |\mathcal{U}| + |\mathcal{I}|$. The user-item interactions are denoted by $\textbf{R} \in \{0,1\}^{|\mathcal{U}| \times |\mathcal{I}|}$, where 0 signifies no interaction, and 1 indicates the presence of an interaction between a user-item pair. To enhance the memory efficiency of the embedding layer, a common approach involves utilizing a compact meta-embedding codebook, i.e., $\textbf{E}_{meta} \in \mathbb{R}^{m \times d}$, with $d$ representing the dimensionality of the embeddings. This codebook serves as a replacement for the more extensive full embedding matrix $\textbf{E} \in \mathbb{R}^{N \times d}$, where the count of meta-embeddings $m$ is significantly smaller than $N$. To generate an embedding for each entity, a common approach involves using an assignment matrix $\textbf{S} \in \mathbb{R}^{N \times m}_{\geq 0}$. Each row of this matrix corresponds to a user or item entity, where the non-zero elements specify which meta-embeddings are selected from the $m$ available options, along with their respective weights in forming the entity's embedding. Consequently, the compositional embeddings for all entities, denoted as $\widehat{\textbf{E}} \in \mathbb{R}^{N \times d}$, can be calculated as follows:
\begin{equation}
\label{eq:cce}
\begin{split}
    \widehat{\textbf{E}} = \textbf{S}\textbf{E}_{meta}.
  \end{split}
\end{equation}
Based on the above, recent developments have demonstrated that Graph Neural Networks (GNNs) can effectively propagate collaborative signals among interconnected entities. Leveraging this, GNN-based recommenders construct a user-item interaction graph to encapsulate the semantic correlations present among entities, as evidenced in their interactions.

\subsection{Graph-Propagated Meta-Embeddings}\label{sec:me}
The GNN-based recommenders typically adopt an adjacency matrix $\textbf{A} \in \mathbb{R}^{N \times N}$ to represent those connections in the user-item interaction graph, facilitating the propagation of collaborative signals among connected entities for embedding learning. The adjacency matrix $\textbf{A}$ is structured as follows:
\begin{equation}
\textbf{A} = \begin{bmatrix}
0 & \textbf{R} \\
\textbf{R}^{T} & 0
\end{bmatrix},
\end{equation}
where $\textbf{R}$ represents the user-item interaction matrix. For each user $u \in \mathcal{U}$ and item $i \in \mathcal{I}$, their embeddings $\textbf{e}_{u}$ and $\textbf{e}_{i}$ are retrieved from $\widehat{\textbf{E}}$ and serve as inputs to the user-item interaction graph $\textbf{A}$. To enhance adaptability, a compositional embedding approach that considers semantic relationships among entities and provides a more refined, weighted combination of assigned meta-embeddings for each entity is proposed. As described in~\cite{LightweightEmbeddings}, meta-embeddings are represented as additional virtual nodes connected to a set of nodes within the user-item interaction graph $\textbf{A}$. The assignment matrix $\textbf{S}$ represents the connections between entities and the meta-embedding codebook. Integrating $\textbf{S}$ into $\textbf{A}$ results in an expanded interaction graph $\textbf{A}'$:
\begin{equation}
\textbf{A}' = \begin{bmatrix}
\textbf{A} & \textbf{S} \\
\textbf{S}^{T} & 0
\end{bmatrix},
\end{equation}
where $\textbf{A}' \in \mathbb{R}^{(N+m) \times (N+m)}$. The input embeddings (layer-0 embeddings) to the GNN are generated by stacking the compositional embedding $\widehat{\textbf{E}}$ and the meta-embedding codebook $\textbf{E}_{meta}$:
\begin{equation}
\textbf{H}_{(0)} = \begin{bmatrix}
\widehat{\textbf{E}} \\
\textbf{E}_{meta}
\end{bmatrix},
\end{equation}
with $\textbf{H}_{(0)} \in \mathbb{R}^{(N+m) \times d}$. The propagation operation at layer $l+1$ is defined as:
\begin{equation}
\textbf{H}_{(l+1)} = (\textbf{D}^{-\frac{1}{2}}\textbf{A}' \textbf{D}^{-\frac{1}{2}})\textbf{H}_{(l)},
\end{equation}
where $\textbf{A}'$ is the adjacency matrix of the expanded interaction graph and $\textbf{D} \in \mathbb{R}^{(N+m) \times (N+m)}$ is the diagonal degree matrix of $\textbf{A}'$. $\textbf{D}^{-\frac{1}{2}}\textbf{A}' \textbf{D}^{-\frac{1}{2}}$ creates a symmetrically normalized adjacency matrix. The final GNN embeddings are obtained by averaging the embeddings from all layers:
\begin{equation}
\textbf{H} = \frac{1}{L+1}\sum_{l=0}^{L}\textbf{H}_{(l)},
\end{equation}
allowing the graph-propagated representations of the full entity set $\textbf{H}_{full} \in \mathbb{R}^{N \times d}$ and the meta-embedding codebook $\textbf{H}_{meta} \in \mathbb{R}^{m \times d}$ to be retrieved by splitting $\textbf{H}$ as follows:
\begin{equation}\label{eq:gpe}
\textbf{H}_{full} \leftarrow \textbf{H}[:N,:], \quad \textbf{H}_{meta} \leftarrow \textbf{H}[N:,:].
\end{equation}
The meta-embeddings $\textbf{H}_{meta}$ incorporate collaborative information from semantically similar user/item neighbors, enabling a more effective meta-embedding assignment for each user/item. For the recommendation task, the user's graph-propagated embedding $\textbf{h}_{u}$ and the item's graph-propagated embedding $\textbf{h}_{i}$ are retrieved from $\textbf{H}_{full}$, and their recommendation affinity prediction is calculated as:
\begin{equation}\label{eq:sim}
\widehat{y}_{ui} = (\textbf{h}_{u})^{T} \textbf{h}_{i}.
\end{equation}
This prediction supports the Bayesian personalized ranking loss (BPR)~\cite{BPR}, which is used to optimize each entity's associated meta-embeddings:
\begin{equation} \label{eq:bpr_loss}
    \mathcal{L}_{\textit{BPR}} = \sum_{(u, i^+, i^-) \in \mathcal{B}} - \ln{\sigma(\widehat{y}_{ui^+}  - \widehat{y}_{ui^-})} + \lambda ||\Theta||^2,
\end{equation}

where $\mathcal{B}$ refers to either the entire training set or a training batch, $(u, i^+, i^-)$ is a triplet consisting of the sampled user $u$'s observed interacted item $i^+$ and unvisited item $i^-$, $||\Theta||^2$ represents the $L_2$ regularization term applied to the trainable parameters, and $\lambda$ determines the regularization weight in the loss function. With a fixed assignment $\mathbf{S}$ in each iteration, the expanded interaction graph $\mathbf{A}'$ utilizes the Eq.\ref{eq:bpr_loss} to back-propagation to update the values in the meta-embedding codebook $\mathbf{E}_{meta}$.

\subsection{Learning Assignment Weights}\label{sec:law}
As the meta-embedding codebook $\textbf{E}_{meta}$ is learned, it is essential to continually update the weights in $\textbf{S}$ to achieve a more accurate mapping between entities and their meta-embedding assignments. For instance, two closely related entities should have their embeddings composed of a similar set of meta-embeddings, reflecting the core principle of collaborative filtering. However, jointly updating the $\textbf{S}$ and $\textbf{E}_{meta}$ will introduce difficulties in finding a stable solution. To solve this challenges,~\cite{LightweightEmbeddings} proposes to preserved the mapping from each entity to its corresponding meta-embeddings after $L$ layers of propagation in the expanded interaction graph $\textbf{A}'$ as same as the  in the input layer of the GNN. Thus:
\begin{equation}\label{eq:law}
\begin{split}
    \textbf{H}_{full} = \textbf{S}\textbf{H}_{meta},
  \end{split}
\end{equation}
where $\textbf{H}_{full} \in \mathbb{R}^{N \times d}$  and $\textbf{H}_{meta} \in \mathbb{R}^{m \times d}$ are the graph-propagated embeddings as in Eq.\ref{eq:gpe}. Based on that, calculating the pseudo-inverse (\textit{i.e.}, $\textbf{H}_{meta}^{\dagger}$) of $\textbf{H}_{meta}$ via the Moore-Penrose inverse~\cite{GeneralizedInverse}, then the assignment matrix $\textbf{S}$ is computed as:
\begin{equation}\label{eq:update_assignment_mat}
\begin{split}
    \textbf{S} = \textbf{H}_{full}\textbf{H}_{meta}^{\dagger}=\textbf{H}_{full}\textbf{V} \Sigma^{-1}\textbf{U}^{*},
  \end{split}
\end{equation}
where $\textbf{V} \in \mathbb{R}^{d \times m}$ is a unitary matrix, $\Sigma^{-1} \in \mathbb{R}^{m \times m}$ is the inverse matrix of  the square matrix $\Sigma$ which contains singular values along the diagonal, $\textbf{U}^{*} \in \mathbb{R}^{m \times m}$ is the conjugate transpose of unitary matrix $\textbf{U} \in \mathbb{R}^{m \times m}$. Updating $\textbf{S}$ via a gradient-free learning strategy enhances computational efficiency. Additionally, this method preserves semantic associations among entities in their meta-embedding assignments. Specifically, similar rows in $\textbf{H}_{\text{full}}$ yield comparable outcomes after the multiplication process in Eq. \ref{eq:update_assignment_mat}.

\section{Overall Framework}
Central to our method lies in three aspects: 1) learning coarse meta-embedding for coarse-grained semantics (Sec.\ref{sec:coarse}); 2) learning sparse fine meta-embedding for fine-grained semantics based on the coarse meta-embedding (Sec.\ref{sec:refine}); 3) how the fine meta-embedding can strength the representation of coarse meta-embedding (Sec.\ref{sec:liaw}). The overall algorithm is in Sec.\ref{sec:algorithm}. Additionally, we provide Table.\ref{symbol}, which explains the descriptions of the various symbols used in the paper.

\subsection{Coarse Graph-Propagated Meta-Embeddings}\label{sec:coarse}
Based on Sec.\ref{sec:me}, we now outline the construction of our coarse graph-propagated meta-embeddings. Initially, we use the coarse meta-embedding codebook $\mathbf{E}_{meta}^c \in \mathbb{R}^{m^c \times d}$ and the coarse assignment matrix $\mathbf{S}^c$ to compute the coarse compositional embeddings $\widehat{\mathbf{E}}^c$ as follows:
\begin{equation}
\begin{split}
    \widehat{\mathbf{E}}^c = \textbf{S}^c\textbf{E}_{meta}^c.
\end{split}
\end{equation}
We then incorporate $\mathbf{S}^c$ into the user-item interaction graph $\mathbf{A}$, resulting in the coarse interaction graph $\mathbf{A}^c \in \mathbb{R}^{(N+m^c) \times (N+m^c)}$:
\begin{equation}\label{eq:ac}
\mathbf{A}^c = \begin{bmatrix}
\mathbf{A} & \mathbf{S}^c \\
(\mathbf{S}^c)^{T} & 0
\end{bmatrix}.
\end{equation}
By stacking $\widehat{\mathbf{E}}^c$ and $\mathbf{E}_{meta}^c$ as the input embeddings of $\mathbf{A}^c$, we obtain the graph-propagated coarse representations for the complete entity set $\mathbf{H}_{full}^c \in \mathbb{R}^{N \times d}$ and the coarse meta-embedding codebook $\mathbf{H}_{meta}^c \in \mathbb{R}^{m^c \times d}$. We then extract the graph-propagated embeddings of entities from $\mathbf{H}_{full}^c$ and compute the similarity score between each user-item pair as defined in Eq.\ref{eq:sim}. With a fixed assignment $\mathbf{S}^c$ in each iteration, the predicted scores of a training batch are fed into Eq.\ref{eq:bpr_loss} to compute the BPR loss $\mathcal{L}_{\textit{BPR}}$, facilitating back-propagation to update the coarse meta-embedding codebook $\mathbf{H}_{meta}^c$. Besides, we will update the $\mathbf{S}^c$ as described in Sec.\ref{sec:law}.

\begin{table}[h]
\centering
\caption{Summary of Symbols of our coarse-to-fine lightweight meta-embedding method} \label{symbol}
\resizebox{0.8\linewidth}{!}{
\begin{tabular}{c|l}
\toprule
\textbf{Symbol} & \textbf{Description} \\
\midrule
$\mathbf{E}_{meta}^c$            & Coarse-grained meta-embedding \\
$\mathbf{E}_{meta}^r$            & Fine-grained meta-embedding \\
$\widehat{\mathbf{E}}_{meta}^c$  & Refined coarse-grained meta-embedding \\
$\widehat{\mathbf{E}}^c$         & Coarse-grained compositional embedding \\
$\widehat{\mathbf{E}}^r$         & Fine-grained compositional embedding \\ \hline
$\mathbf{R}$                     & User-item interaction matrix \\
$\mathbf{S}^c$                   & Coarse-grained assignment matrix \\
$\mathbf{S}^r$                   & Coarse-to-fine assignment matrix \\
$\mathbf{A}$                     & Adjacency matrix for user-item interactions \\
$\mathbf{A}^c$                   & Coarse-grained interaction graph \\
$\mathbf{A}^r$                   & Fine-grained interaction graph \\  \hline
$\mathbf{H}_{\text{meta}}^c$     & Coarse-grained meta-embedding codebook \\
$\mathbf{H}_{\text{meta}}^r$     & Fine-grained meta-embedding codebook \\
$\mathbf{H}_{\text{full}}^c$     & Graph-propagated coarse-grained representations for the full entity set \\
$\mathbf{H}_{\text{full}}^r$     & Graph-propagated fine-grained representations for the full entity set \\ \hline
$w_{cr}$                         & Weight factor for coarse-to-fine transition \\
$\mathcal{S}_{\lambda}$          & Soft thresholding function \\
\bottomrule
\end{tabular}}
\end{table}

\subsection{Fine Graph-Propagated Meta-Embeddings}\label{sec:refine}
Based on the coarse meta-embedding, it is crucial to capture more fine representations to improve upon coarse meta-embeddings. To learn more fine-grained semantic information, rather than directly connecting coarse meta-embeddings to a set of entity nodes for representation learning, we propose connecting these fine meta-embeddings as auxiliary nodes to the coarse meta-embedding nodes. This approach prevents learning redundant information from the coarse stage. Consequently, we can construct the fine interaction graph $\mathbf{A}^r \in \mathbb{R}^{(N + m^c + m^r) \times (N + m^c + m^r)}$ as:
\begin{equation}\label{eq:ar}
\mathbf{A}^r = \begin{bmatrix}
\mathbf{A} &\mathbf{S}^{c} &0 \\
(\mathbf{S}^{c})^{T} &0 &\mathbf{S}^{r} \\
0 &(\mathbf{S}^{r})^{T} &0
\end{bmatrix},
\end{equation}
where $\mathbf{S}^{r} \in \mathbb{R}^{m^c \times m^r}$ is the indirect assignment matrix between the two kinds of meta-embeddings. Before discussing how to refine the coarse meta-embedding codebook, we need to learn the fine-grained representation $\mathbf{E}_{meta}^r \in \mathbb{R}^{m^r \times d}$.

\begin{figure*}
  \centering
  \includegraphics[width=\textwidth]{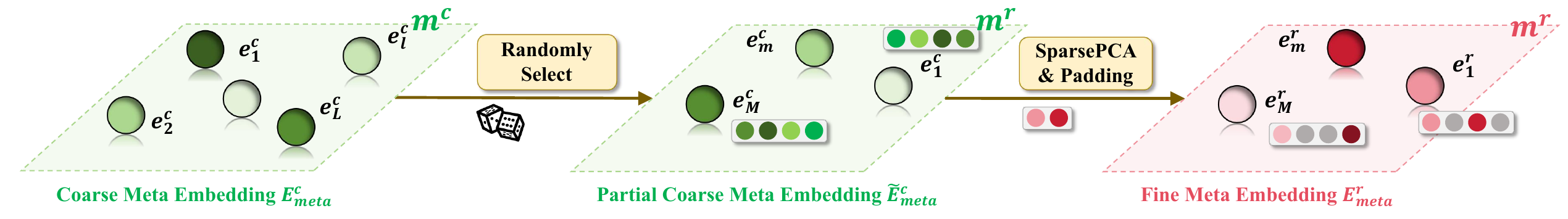}
  \caption{Illustration of fine meta embedding initialization strategy, which preserves sparsity and maintains associative relationships between users and items}\label{fig:init}
\end{figure*}

As shown in Fig.\ref{fig:init}, we propose a method to improve the initialization process for the coarse meta-embedding codebook $\mathbf{E}_{meta}^c$. Instead of randomly assigning values, we suggest selecting $r$ coarse meta-embeddings from $\mathbf{E}_{meta}^c$ to form a new matrix $\mathbf{\widetilde{E}}_{meta}^c$. This selection process allows us to initialize $\mathbf{E}_{meta}^r$ based on these chosen coarse meta-embeddings, thus enhancing the semantic relationships between entities from the start. Furthermore, when refining the coarse meta-embeddings, we aim to avoid learning dense vectors identical to the coarse ones, which would result in a dense matrix $\mathbf{E}_{meta}^r$. Instead, our objective is to make $\mathbf{E}_{meta}^r$ sparse. This approach serves two purposes. Firstly, it ensures that the coarse stage focuses on capturing coarse-grained semantic information, while the fine stage deals with representing fine-grained details as auxiliary information. Secondly, it aligns with the goal of compressing the total number of meta-embeddings across both stages. To achieve more flexible initialization of sparse $\mathbf{E}_{meta}^r$, we propose utilizing SparsePCA (Principal Component Analysis) for initialization. SparsePCA inherits PCA's benefits in effectively capturing the main patterns of variability while reducing computational complexity and feature redundancy. Moreover, it introduces the capability to control sparsity within the extracted components. This control over sparsity is crucial as it allows us to regulate the degree of sparsity in the meta-embedding matrix. By leveraging SparsePCA, we can tailor the initialization process to better induce sparsity in the meta-embedding matrix, thereby enhancing the overall efficiency and effectiveness of the learning process. We have:
\begin{equation}\label{spacepca}
\min_{\mathbf{\widetilde{E}}_{meta}^r, W^r} \frac{1}{2} \|\mathbf{\widetilde{E}}_{meta}^c - \mathbf{\widetilde{E}}_{meta}^r W^r\|_F^2 + \alpha \|W^r\|_1,
\end{equation}
where $\|\cdot\|_F^2$ represents the reconstruction error (Frobenius norm), $\|\cdot\|_1$ denotes the sparsity penalty term, and $\alpha$ is the hyperparameter controlling sparsity. The objective of SparsePCA is to find a $d_r$ dimensional sparse representation $\widetilde{E}_{meta}^r \in \mathbb{R}^{m^r \times d^r}$ along with the corresponding sparse coefficient matrix $W^r \in \mathbb{R}^{d^r \times d}$. Subsequently, the sparse vectors $\mathbf{\widetilde{E}}_{meta}^r$ resulting from SparsePCA dimensionality reduction are padded with zeros to revert them to the specified $d$-dimensional vectors $\mathbf{E}_{meta}^r \in \mathbb{R}^{m^r \times d}$.

However, SparsePCA, while effective in producing sparse representations, suffers from non-differentiability issues that prevent maintaining sparsity in embeddings during training. To address this, we propose the use of soft thresholding, defined as:
\begin{equation}\label{eq:softthresholding}
\mathcal{S}_{\lambda}(\mathbf{E}_{meta}^r) = \text{sign}(\mathbf{E}_{meta}^r) \cdot \max(|\mathbf{E}_{meta}^r| - \lambda, 0),
\end{equation}
where $\text{sign}(x)$ is the sign function, which returns $-1$ if $x < 0$, $1$ if $x > 0$, and $0$ if $x = 0$, $|\cdot|$ denotes the absolute value, and $\lambda$ is the threshold parameter. The soft thresholding technique is integrated to dynamically adjust the sparsity levels during training, providing a more adaptable and robust model that responds to the varying intricacies of the data by smoothly driving small coefficients to zero while retaining the significant ones. This method ensures that the resulting embeddings remain sparse and computationally efficient, facilitating better convergence in training processes. Thanks to $\mathbf{E}_{meta}^r$, the coarse compositional embeddings $\widehat{\mathbf{E}}^c$ in Eq.\ref{eq:cce} can be reformulated as follows:
\begin{align}
& \widehat{\mathbf{E}}_{meta}^c = (1 - w_{cr}) \mathbf{E}_{meta}^c + w_{cr} \mathbf{S}^r \mathcal{S}_{\lambda}(\mathbf{E}_{meta}^r), \\
& \widehat{\mathbf{E}}^r = \mathbf{S}^c \widehat{\mathbf{E}}_{meta}^c,
\end{align}
where $\widehat{\mathbf{E}}^r \in \mathbb{R}^{N \times d}$ is the refined compositional embeddings, and $w_{cr}$ is the weight parameter. The input embeddings to the refined GNN are stacking the $\widehat{\mathbf{E}}^c$, $\widehat{\mathbf{E}}_{meta}^c$ and $\mathcal{S}_{\lambda}(\mathbf{E}_{meta}^r)$:
\begin{equation}
\mathbf{H}_{(0)}^{r} = \begin{bmatrix}
\widehat{\mathbf{E}}^c \\
\widehat{\mathbf{E}}_{meta}^c \\
\mathcal{S}_{\lambda}(\mathbf{E}_{meta}^r)
\end{bmatrix},
\end{equation}
where $\mathbf{H}_{(0)}^{r} \in \mathbb{R}^{(N + m^c + m^r) \times d}$, following the adjacency matrix of the refined interaction graph $\mathbf{A}^r$, the propagation operation at layer $l+1$ is:
\begin{equation}
\mathbf{H}^r_{(l+1)} = (\widetilde{\mathbf{D}}^{-\frac{1}{2}} \mathbf{A}^r \widetilde{\mathbf{D}}^{-\frac{1}{2}}) \mathbf{H}^r_{(l)},
\end{equation}
where $\widetilde{\mathbf{D}} \in \mathbb{R}^{(N + m^c + m^r) \times (N + m^c + m^r)}$ is the diagonal degree matrix of $\mathbf{A}^r$. The final GNN embeddings are obtained by averaging the embeddings from all layers:
\begin{equation}
\mathbf{H}^r = \frac{1}{L+1} \sum_{l=0}^{L} \mathbf{H}_{(l)}^r,
\end{equation}
allowing the graph-propagated representations of the full entity set $\mathbf{H}^r_{full} \in \mathbb{R}^{N \times d}$, the coarse meta-embedding codebook $\widehat{\mathbf{H}}_{meta}^c \in \mathbb{R}^{m^c \times d}$, and the fine meta-embedding codebook ${\mathbf{H}}_{meta}^r \in \mathbb{R}^{m^r \times d}$ to be retrieved by splitting $\mathbf{H}^r$ as follows:
\begin{equation}\label{eq:repro}
\mathbf{H}^r_{full} \leftarrow \mathbf{H}^r[:N,:], \quad \widehat{\mathbf{H}}_{meta}^c \leftarrow \mathbf{H}^r[N:N+m^c,:], \quad {\mathbf{H}}_{meta}^r \leftarrow \mathbf{H}^r[N+m^c:,:].
\end{equation}
We then extract the graph-propagated embeddings of entities from $\mathbf{H}_{full}^r$ and compute the similarity score between each user-item pair as defined in Eq.\ref{eq:sim}. With a fixed assignment $\mathbf{S}^c$, $\mathbf{S}^r$, and $\mathbf{E}_{meta}^c$ in each iteration, the predicted scores of a training batch are fed into Eq.\ref{eq:bpr_loss} to compute the BPR loss $\mathcal{L}_{\textit{BPR}}$, facilitating back-propagation to update the fine meta-embedding codebook $\mathbf{H}_{meta}^r$.

\begin{figure*}
  \centering
  \includegraphics[width=0.85\textwidth]{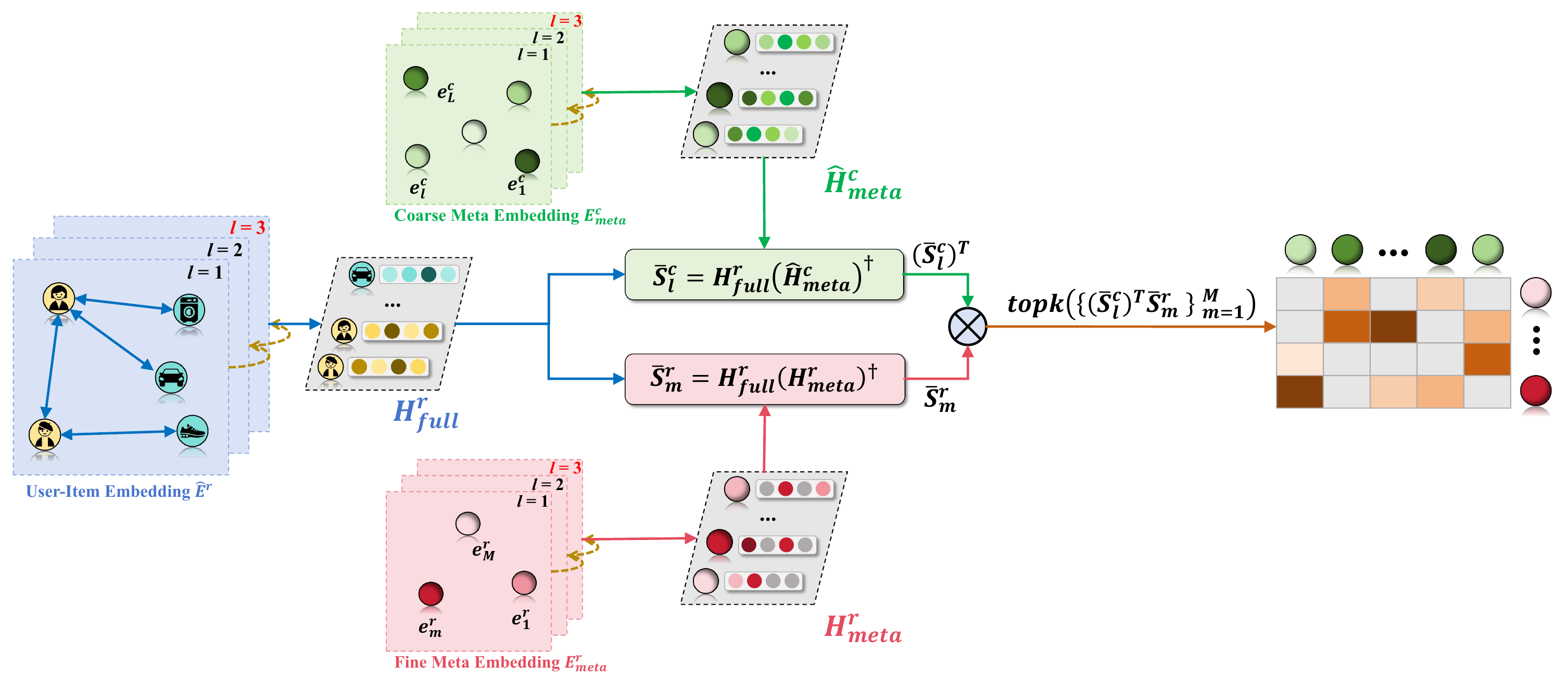}
  \caption{Illustration of weight bridging updating strategy for $S^{r}$, which dynamically aligns coarse meta-embeddings with multiple fine meta-embeddings based on users/items' semantic relevance.}\label{gate}
\end{figure*}
\subsection{Learning Coarse-to-Fine Assignment Weights} \label{sec:liaw}
Unlike the coarse stage, which can directly calculate the value of $\mathbf{S}^c$ using Eq.\ref{eq:update_assignment_mat}, the fine interaction graph is constructed hierarchically. Therefore, we need to calculate the weight matrix $\mathbf{S}^r$ between two types of meta-embedding codebooks. We aim for each coarse meta-embedding to efficiently select the corresponding fine meta-embedding as a semantic refinement. However, directly calculating the similarity score between coarse and fine meta-embedding would ignore the entities' semantics.
As mentioned earlier, Eq.\ref{eq:law} shows that the mapping relationship from each entity to its corresponding meta-embeddings, after L layers of propagation, remains the same as in the input layer of the GNN. In this spirit, entities can serve as a bridge~\cite{DGTS}, linking each coarse meta-embedding with several refined meta-embeddings according to their semantics. Thus, we propose the following bridging updating strategy for the coarse-to-fine assignment matrix (as illustrated in Fig.\ref{gate}):
\begin{equation}\label{eq:liaw}
\begin{split}
\mathbf{S}^r = (\textbf{H}^r_{full}(\widehat{\textbf{H}}^c_{meta})^{\dagger})^T(\textbf{H}^r_{full}(\textbf{H}^r_{meta})^{\dagger}),
\end{split}
\end{equation}
where $(\widehat{\textbf{H}}^c_{meta})^{\dagger}$ is the pseudo-inverse of $\widehat{\textbf{H}}^c_{meta}$, and $(\textbf{H}^r_{meta})^{\dagger}$ is the pseudo-inverse of $\textbf{H}^r_{meta}$, both calculated using the Moore-Penrose inverse. Since Eq.\ref{eq:liaw} yields a dense matrix, to maintain the sparsity of $\mathbf{S}^r$, we perform sparsification by retaining the largest $t^r$ non-zero weights in each row of $\mathbf{S}^r$. This ensures that each coarse meta-embedding uses exactly $t^r$ fine meta-embeddings at all times:
\begin{equation}\label{ar_sparse}
\mathbf{S}^r[p,q] = \begin{cases}
\mathbf{S}^r[p,q] & \text{if}, q \in \text{idx}_{\text{top-}(t^r)}(\mathbf{S}^r[p,:]) \\
0 & \text{otherwise}
\end{cases},
\end{equation}
where $\text{idx}_{\text{top-}(t^r)}(\cdot)$ returns the indices of the top-$(t^r)$ entries of a given vector. Essentially, we retain the relatively larger weights in $\mathbf{S}^r$ only for those fine meta-embeddings considered critical for a coarse meta-embedding. Last but not least, given the fine-grained meta-embedding, a well-initialized coarse-to-fine assignment matrix \(S^r\) can significantly contribute to improving the recommendation accuracy of the coarse-grained meta-embedding and accelerating training. The goal is to link each coarse meta-embedding with several refined meta-embeddings based on their semantics. To achieve this, we leverage the METIS~\cite{METIS} algorithm, a well-established multilevel graph partitioning method, to partition the user-item interaction graph. While other advanced graph clustering techniques may be viable options, METIS stands out for this one-time initialization because it offers fast and accurate computation without requiring any learning. Additionally, it provides balanced and non-overlapping partitions that ensure fair utilization of all anchor meta-embeddings, along with deterministic results that facilitate easy replication. To reflect semantics in the initialization of the assignment matrix \(S^r\), we perform the following for each fine-grained meta-embedding:

1. Given entity \(p\)'s subgraph index \(c_p \in \{1, 2, \ldots, c\}\), we set \(\tilde{S}^r[p, q^{*}_p] = w^*\) and \(\tilde{S}^r[p, q] = \frac{1-w^*}{c-1}\), where \(q^{*}_p = c_p\), and \(q \in \{1, 2, \ldots, c\} \setminus c_p\), with \(w^*\) being a universal hyperparameter and \(\tilde{S}^r \in \mathbb{R}^{N \times m^r}\).

2. Using a weight-bridging update strategy, we calculate \(\hat{S}^r = (S^c)^T(\tilde{S}^r)\), enabling entities to act as bridges linking each coarse meta-embedding with several refined meta-embeddings according to their semantics.

3. For sparsity, we retain only the relatively larger weights in \(\hat{S}^r\) to form \(S^r\), ensuring that only the fine-grained meta-embeddings critical for an entity's embedding are considered .

\subsection{The Overall Algorithm}\label{sec:algorithm}
\noindent The provided pseudocode outlines the two stages of our algorithm: the Coarse Stage (Algorithm \ref{algorithm}) and the Refined Stage (Algorithm \ref{algorithm1}). In the coarse stage, we initialize the coarse meta-embedding matrix $\textbf{E}^c_{meta}$ and the assignment matrix $\textbf{S}^c$. The update frequency for this stage, denoted as $f^c$, is set to control the frequency of assignment updates. Additionally, we compute the adjacency matrix $\mathbf{A}^c$ based on a specific equation. During iterative optimization of the Bayesian Personalized Ranking ($\mathcal{L}_{BPR}$) loss, we periodically update the assignment matrix and regenerate $\mathbf{A}^c$ to reflect any changes in entity-meta-embedding associations. Transitioning to the fine stage, we initialize the fine meta-embedding matrix $\textbf{E}^r_{meta}$ and the coarse-to-fine assignment matrix $\textbf{S}^r$ via our proposed strategy. Here, we freeze the coarse meta-embedding matrix $\textbf{E}^c_{meta}$. Similar to the coarse stage, we determine the update frequency $f^r$ for this stage. Following initialization, we compute the adjacency matrix $\mathbf{A}^r$ to capture fine relationships among entities and meta-embeddings. During iterative optimization of the $\mathcal{L}_{BPR}$ loss, we update the coarse-to-fine assignment matrix and regenerate $\mathbf{A}^r$ at intervals dictated by $f^r$, ensuring consistent refinement and adaptation to evolving data dynamics.

\begin{algorithm}
    \caption{The Coarse-Grained Training Stage} 
    \label{algorithm}
    \LinesNumbered
    Randomly initialize $\textbf{E}^c_{meta} \in \mathbb{R}^{m^c \times d}$\;
    Initialize $\textbf{S}^c \in \mathbb{R}^{N \times m^c}$\;
    Initialize the update frequency for coarse stage $f^c$\;
    Compute $\mathbf{A}^c \in \mathbb{R}^{(N+m^c) \times (N+m^c)}$ with Eq.\ref{eq:ac}\;

    \While{not converged}{
        Optimize $\mathcal{L}_{BPR}$ following Sec.\ref{sec:coarse}\;
    }
        epoch\_index $\gets 0$\;

        \While{not converged}{
            Optimize $\mathcal{L}_{BPR}$ following Sec.\ref{sec:coarse}\;
            \If{epoch\_index $\textit{mod } f^c == 0$}{
                Perform assignment update with Eq.\ref{eq:update_assignment_mat}\;
                Regenerate $\mathbf{A}^c$ with Eq.\ref{eq:ac}\;
            }
            epoch\_index $\gets$ epoch\_index $+ 1$\;
        }
\end{algorithm}

\begin{algorithm}
    \caption{The Fine-Grained Training Stage} 
    \label{algorithm1}
    \LinesNumbered
    Initialize $\textbf{E}^r_{meta} \in \mathbb{R}^{m^r \times d}$ with Eq.\ref{spacepca}\;
    Initialize $\textbf{S}^r \in \mathbb{R}^{m^c \times m^r}$\;
    Initialize the update frequency for refined stage $f^r$\;
    Freeze the $\textbf{E}^c_{meta}$ and $\textbf{S}^c$\;
    Compute $\mathbf{A}^r \in \mathbb{R}^{(N+m^c+m^r) \times (N+m^c+m^r)}$ with Eq.\ref{eq:ar}\;

    \While{not converged}{
        Optimize $\mathcal{L}_{BPR}$ following Eq.\ref{eq:softthresholding}--\ref{eq:repro}\;
    }
        epoch\_index $\gets 0$\;
        \While{not converged}{
                Optimize $\mathcal{L}_{BPR}$ following Eq.\ref{eq:softthresholding}--\ref{eq:repro}\;
            \If{epoch\_index $\textit{mod } f^r == 0$}{
                Perform assignment update with Eq.\ref{eq:liaw} and Eq.\ref{ar_sparse}\;
                Regenerate $\mathbf{A}^r$ with Eq.\ref{eq:ar}\;
            }
            epoch\_index $\gets$ epoch\_index $+ 1$\;
        }
\end{algorithm}

\begin{table}
    \caption{Statistics of datasets used in our work.}
    \label{tab:dataset_stat}
    \centering
    \resizebox{0.5\textwidth}{!}{
    \begin{tabular}{c | c c  c c}
        \toprule
        \textbf{Dataset} & \#\textbf{User} & \#\textbf{Item} & \#\textbf{Interactions} & \textbf{Density}\\
        \midrule
        \textbf{Gowalla} & 29,858 & 40,981 & 1,027,370 & 0.084\%\\
        \textbf{Yelp2020}  & 71,135  & 45,063 & 1,782,999 & 0.056\%\\
        \textbf{Amazon-book} & 52,643 & 91,599 & 2,984,108 & 0.062\%\\
        \bottomrule
    \end{tabular}
    }
\end{table}
\section{Experiments}
In this section, we conduct a series of experiments to validate the effectiveness of our work. We organize our experiments around the following research questions (RQs):
\begin{itemize}
    \item[$\bullet$] \textbf{Q1:} How is the recommendation accuracy of our work compared to other baselines?
    \item[$\bullet$] \textbf{Q2:} What is the effect of the key components?
    \item[$\bullet$] \textbf{Q3:} How sensitive is our work to hyperparameters?
    \item[$\bullet$] \textbf{Q4:} How do fine meta-embeddings enhance lightweight implementation?
\end{itemize}
\subsection{Experimental Settings}
\subsubsection{Datasets.} In our evaluation, we engage three distinct benchmark data corpus: Gowalla~\cite{gowalla}, Yelp2020~\cite{Yelp}, and Amazon-book~\cite{amazon-book}. They are notable for their open accessibility and the diversity they encompass in scale, sectors, and user-item interaction density. We split the train/test/validation dataset according to~\cite{GeneralizedInverse}. The details of these datasets are inventoried in Table \ref{tab:dataset_stat}.
\begin{itemize}
    \item[(1)] \textbf{Gowalla}: Originating from the social networking platform Gowalla, this dataset captures user-generated location check-ins.
    \item[(2)] \textbf{Yelp2020}: This dataset is procured from Yelp's 2020 challenge data trove, encompassing entities such as local service providers, restaurants, and bars.
    \item[(3)] \textbf{Amazon-Book}: This subset is extracted from the Amazon review ecosystem, a prevalent benchmark in product recommendation research, specifically focusing on books.
\end{itemize}
\subsubsection{Baseline Algorithms.} To validate the superiority of our method, we compare it with the typical recommender models, including: \textbf{PEP}~\cite{PEP}, \textbf{QR}~\cite{QR} and \textbf{LEGCF}~\cite{LightweightEmbeddings} allows the generation of embedding layers to satisfy various memory targets; \textbf{AutoEmb}~\cite{AutoEmb}, \textbf{ESAPN}~\cite{ESAPN}, \textbf{OptEmbed}~\cite{OptEmbed}, \textbf{CIESS}~\cite{CIESS}, \textbf{DHE}~\cite{DHE}, \textbf{NimbleTT}~\cite{NimbleTT} ignore the final embedding structure meets the memory budget, because these methods do not formalize the memory target as an optimization objective; we also provide the results of
the non-compressed versions of \textbf{LightGCN} with relatively large dimension sizes, \textit{i.e.}, \textbf{UD - dim 128} and \textbf{UD - dim 64}, to compare the performance compromise of various lightweight embedding methods. Due to those on-device recommenders are designed for session-based recommendation, so we can not provide their results of id-based recommendation.

\subsubsection{Evaluation Metrics.} Following metrics are used to evaluate compared methods, we set N to values of \{5, 10, 20\} to observe recommendations' performance.
\begin{itemize}
    \item[(1)] \textbf{NDCG@N} (Normalized Discounted Cumulative Gain): This metric appraises the ranking quality by gauging the graded relevance of the recommended items. Unlike precision-based metrics, NDCG@N accounts for the decay of relevance as users are less likely to consider items lower in their feed. It assigns maximum gain to the top-ranked relevant recommendations, attenuating the value logarithmically as the rank diminishes.
    \item[(2)] \textbf{Recall@N}: Embodies the proportion of a user's preferred items that appear in the top-N slice of the recommendation list. Recall@N is insightful for discerning the coverage breadth of a recommender system, with an ideal system uncovering all items of interest in the user's top-tier selections.
\end{itemize}
\begin{table*}[t]
    \centering
    \caption{Performance comparison between our method and baselines. ``\#Param''  indicates the total parameter size of the \underline{embedding layer} of each method. ``UD - dim 128'' and ``UD - dim 64'' are the full embedding table setting with unified dimensions $d=128$ and $d=64$, respectively. In each column, we use \textbf{bold font} to mark the best result achieved by a lightweight embedding method.}
    \label{tab:qr}
    \resizebox{\textwidth}{!}{
    \begin{tabular}{@{}cc ccccc ccccc ccccc@{}}
    \toprule
	\multirow{2}{*}{\textbf{Method}}  &\multirow{2}{*}{\textbf{Venue}}
    & \multicolumn{5}{c }{\textbf{Gowalla}}  &\multicolumn{5}{c }{\textbf{Yelp2020}}& \multicolumn{5}{c}{\textbf{Amazon-book}}\\ \cmidrule(lr){3-7} \cmidrule(lr){8-12} \cmidrule(l){13-17}
     & &\textbf{\#Param} & \textbf{N@10} & \textbf{R@10} & \textbf{N@20} & \textbf{R@20} &\textbf{\#Param} & \textbf{N@10} & \textbf{R@10} & \textbf{N@20} & \textbf{R@20} &\textbf{\#Param} & \textbf{N@10} & \textbf{R@10} & \textbf{N@20} & \textbf{R@20} \\ \cmidrule(r){1-2} \cmidrule(lr){3-3} \cmidrule(lr){4-5} \cmidrule(lr){6-7} \cmidrule(lr){8-8} \cmidrule(lr){9-10} \cmidrule(lr){11-12} \cmidrule(lr){13-13} \cmidrule(lr){14-15} \cmidrule(l){16-17}
        UD - dim128  &- & 9.07m & 0.0901 & 0.1101 & 0.1059 & 0.1576 & 14.87m & 0.0284 & 0.0426 & 0.0382 &0.0721 &18.46m &0.0172 &0.0215 &0.0230 &0.0367\\
        UD - dim 64   &-  & 4.53m & 0.0884 & 0.1088 & 0.1041 & 0.1557 & 7.44m & 0.0274 & 0.0409 & 0.0366 &0.0687 &9.23m &0.0165 &0.0204 &0.0222 &0.0353\\
        \midrule
        ESAPN  &SIGIR '20  & 0.78m & 0.0268 & 0.0273 & 0.0305 & 0.0405 & 1.12m & 0.0066 & 0.0083 & 0.0087 &0.0141 &1.44m &0.0045 &0.0037 &0.0051 &0.0057\\
        AutoEmb &ICDM '21 & 0.92m & 0.0273 & 0.0278 & 0.0319 & 0.0429 & 1.51m & 0.0071 & 0.0089 & 0.0093 &0.0148 &1.44m &0.0048 &0.0039 &0.0059 &0.0070\\
        DHE &SIGKDD '21 & 0.85m & 0.0049 & 0.0051 & 0.0068 & 0.0105 & 1.39m & 0.0020 & 0.0025 & 0.0028 &0.0046 &1.73m &0.0012 &0.0010 &0.0015 &0.0019\\
        OptEmbed &CIKM '22 & 1.42m & 0.0375 & 0.0370 & 0.0405 & 0.0498 & 5.96m & 0.0085 & 0.0090 & 0.0109  &0.0159 &2.11m &0.0038 &0.0032 &0.0050 &0.0062\\
        NimbleTT &SIGKDD '22 & 0.20m & 0.0259 & 0.0296 & 0.0300 & 0.0431 & 0.28m & 0.0055 & 0.0073 & 0.0071  &0.0122 &0.31m &0.0028 &0.0031 &0.0037 &0.0055\\
        CIESS &SIGIR '23 & 0.29m & 0.0643 & 0.0742 & 0.0745 & 0.1080 & 0.47m & 0.0175 & 0.0253 & 0.0234  &0.0437 &0.57m &0.0021 &0.0030 &0.0029 &0.0054\\
        \midrule
        QR &SIGKDD '20 & 0.21m & 0.0354 & 0.0416 & 0.0412 & 0.0606 & 0.30m & 0.0067 & 0.0092 & 0.0088  &0.0157 &0.29m &0.0040 &0.0046 &0.0053 &0.0082\\
        PEP &ICLR '21 & 0.21m & 0.0638 & 0.0568 & 0.0698 & 0.0807 & 0.30m & 0.0170 & 0.0191 & 0.0218 &0.0326 &0.29m &0.0030 &0.0024 &0.0037 &0.0043\\
        LEGCF &SIGIR '24 & 0.21m & 0.0846 & 0.0979 & 0.0988 & \textbf{0.1444} & 0.30m & 0.0214 & 0.0310 & 0.0291 &0.0548 &0.35m &0.0134 &0.0156 &0.0172 &0.0259\\
        \midrule
        \textbf{Ours} &- &0.19m &\textbf{0.0880} &\textbf{0.0997} &\textbf{0.1009} &0.1430 &0.29m &\textbf{0.0223} &\textbf{0.0323} &\textbf{0.0299} &\textbf{0.0555}  &0.34m & \textbf{0.0146}&\textbf{0.0165} &\textbf{0.0188} &\textbf{0.0280}  \\
    \bottomrule
    \end{tabular}
  \label{table:compare1}
  }
\end{table*}

\subsubsection{Implementation Details.}
In our work, we set the embedding size \( d \) for meta-embedding to 128. To create the training set, we randomly draw 5 negative items for each user-positive item interaction. Meta-embeddings and recommender weights are initialized using Xavier Initialization~\cite{UnderstandDNN}, and the ADAM optimizer~\cite{Adam} is employed for training.
In the coarse training stage, to ensure that the model learns a better assignment matrix, we update the coarse assignment matrix every epoch and set the early stop threshold to 10 for both the fixed assignment training stage and the assignment update stage. The learning rate is set to \(10^{-3}\), and the L2 penalty factor \(\lambda\) is \(5 \times 10^{-4}\). In contrast, for the fine-grained training stage, we maintain the same update frequency for the coarse-to-fine assignment matrix as in the coarse assignment matrix. However, to reduce complexity, we set the early stop threshold to 5 for both the fixed assignment training stage and the assignment update stage. The learning rate for this stage is \(3 \times 10^{-3}\), with the L2 penalty factor \(\lambda\) also set to \(5 \times 10^{-4}\). We utilize a 3-layer GNN for the Gowalla dataset and a 4-layer GNN for the Yelp2020 and Amazon-book datasets. The codebook is divided into two parts: 300 buckets for coarse meta-embeddings and 100 buckets for fine meta-embeddings. The overall performance evaluated with this configuration is discussed in RQ1. For variable-size baselines, we first calculate the total number of parameters used in our 500-bucket configuration. This memory budget is then applied to set the appropriate sparsity target or number of hashing buckets for methods like PEP and QR, ensuring that their embedding layers do not exceed our specified memory constraints. For baselines that lack control over final parameter sizes, we establish suitable search spaces to avoid excessive memory usage while providing flexibility for optimal setting selection. All the experiments are implemented with the pytorch framework and run on NVIDIA A6000.

\begin{table}[h]
\centering
\caption{NDCG and Recall values at different cutoffs (@1, @5, @10, @20) with p-values and improvements, \textcolor{red}{$\uparrow$} denotes improvement of NDCG and Recall compared to LEGCF.}\label{p_value}
\resizebox{\linewidth}{!}{
\begin{tabular}{c|llllc|llllc}
\toprule
\textbf{Methods (Seed)} & \textbf{NDCG@1} & \textbf{NDCG@5} & \textbf{NDCG@10} & \textbf{NDCG@20} & \textbf{P-value} & \textbf{Recall@1} & \textbf{Recall@5} & \textbf{Recall@10} & \textbf{Recall@20} & \textbf{P-value} \\
\midrule
\multicolumn{11}{c}{Dataset: Gowalla} \\
\midrule
LEGCF (2024) & 0.0788 & 0.0856 & 0.0984 & 0.1213 & - & 0.0678 & 0.0978 & 0.1405 & 0.2233 & - \\
Ours (2024) & 0.0829 \textcolor{red}{\scriptsize$\uparrow$ 5.20\%} & 0.0897 \textcolor{red}{\scriptsize$\uparrow$ 4.79\%} & 0.1021 \textcolor{red}{\scriptsize$\uparrow$ 3.76\%} & 0.1247 \textcolor{red}{\scriptsize$\uparrow$ 2.80\%} & 0.00025 & 0.0712 \textcolor{red}{\scriptsize$\uparrow$ 5.01\%} & 0.1016 \textcolor{red}{\scriptsize$\uparrow$ 3.88\%} & 0.1433 \textcolor{red}{\scriptsize$\uparrow$ 1.99\%} & 0.2251 \textcolor{red}{\scriptsize$\uparrow$ 0.81\%} & 0.00778 \\
Ours (2025) & 0.0840 \textcolor{red}{\scriptsize$\uparrow$ 6.60\%} & 0.0906 \textcolor{red}{\scriptsize$\uparrow$ 5.84\%} & 0.1032 \textcolor{red}{\scriptsize$\uparrow$ 4.88\%} & 0.1256 \textcolor{red}{\scriptsize$\uparrow$ 3.55\%} & \textbf{0.00015} & 0.0717 \textcolor{red}{\scriptsize$\uparrow$ 5.75\%} & 0.1020 \textcolor{red}{\scriptsize$\uparrow$ 4.29\%} & \textbf{0.1446} \textcolor{red}{\scriptsize$\uparrow$ 2.92\%} & 0.2256 \textcolor{red}{\scriptsize$\uparrow$ 1.03\%} & \textbf{0.00421} \\
Ours (2026) & \textbf{0.0843} \textcolor{red}{\scriptsize$\uparrow$ 7.00\%} & \textbf{0.0907} \textcolor{red}{\scriptsize$\uparrow$ 5.96\%} & \textbf{0.1034} \textcolor{red}{\scriptsize$\uparrow$ 5.08\%} & \textbf{0.1258} \textcolor{red}{\scriptsize$\uparrow$ 3.71\%} & 0.00018 & \textbf{0.0721} \textcolor{red}{\scriptsize$\uparrow$ 6.34\%} & \textbf{0.1022} \textcolor{red}{\scriptsize$\uparrow$ 4.50\%} & \textbf{0.1446} \textcolor{red}{\scriptsize$\uparrow$ 2.92\%} & \textbf{0.2257} \textcolor{red}{\scriptsize$\uparrow$ 1.08\%} & 0.00428 \\
\midrule
\multicolumn{11}{c}{Dataset: Yelp2020} \\
\midrule
LEGCF (2024) & 0.0167 & 0.0214 & 0.0287 & 0.0417 & - & 0.0177 & 0.0306 & 0.0533 & 0.1017 & - \\
Ours (2024) & 0.0175 \textcolor{red}{\scriptsize$\uparrow$ 4.79\%} & 0.0222 \textcolor{red}{\scriptsize$\uparrow$ 3.74\%} & 0.0297 \textcolor{red}{\scriptsize$\uparrow$ 3.48\%} & 0.0429 \textcolor{red}{\scriptsize$\uparrow$ 2.88\%} & \textbf{0.00159} & 0.0185 \textcolor{red}{\scriptsize$\uparrow$ 4.52\%} & 0.0318 \textcolor{red}{\scriptsize$\uparrow$ 3.92\%} & 0.0549 \textcolor{red}{\scriptsize$\uparrow$ 3.00\%} & 0.1043 \textcolor{red}{\scriptsize$\uparrow$ 2.56\%} & 0.02548 \\
Ours (2025) & \textbf{0.0177} \textcolor{red}{\scriptsize$\uparrow$ 5.99\%} & \textbf{0.0225} \textcolor{red}{\scriptsize$\uparrow$ 5.14\%} & \textbf{0.0302} \textcolor{red}{\scriptsize$\uparrow$ 5.23\%} & \textbf{0.0439} \textcolor{red}{\scriptsize$\uparrow$ 5.27\%} & 0.00937 & \textbf{0.0189} \textcolor{red}{\scriptsize$\uparrow$ 6.78\%} & \textbf{0.0322} \textcolor{red}{\scriptsize$\uparrow$ 5.23\%} & \textbf{0.0560} \textcolor{red}{\scriptsize$\uparrow$ 5.07\%} & \textbf{0.1072} \textcolor{red}{\scriptsize$\uparrow$ 5.41\%} & 0.06585 \\
Ours (2026) & 0.0175 \textcolor{red}{\scriptsize$\uparrow$ 4.79\%} & 0.0223 \textcolor{red}{\scriptsize$\uparrow$ 4.21\%} & 0.0297 \textcolor{red}{\scriptsize$\uparrow$ 3.48\%} & 0.0431 \textcolor{red}{\scriptsize$\uparrow$ 3.36\%} & 0.00194 & 0.0187 \textcolor{red}{\scriptsize$\uparrow$ 5.65\%} & 0.0321 \textcolor{red}{\scriptsize$\uparrow$ 4.90\%} & 0.0548 \textcolor{red}{\scriptsize$\uparrow$ 2.81\%} & 0.1045 \textcolor{red}{\scriptsize$\uparrow$ 2.75\%} & \textbf{0.02153} \\
\bottomrule
\end{tabular}}
\end{table}
\subsection{Overall Performance (Q1)}
We set the coarse meta-embedding bucket size \(m^c=300\) and the fine meta-embedding bucket size \(m^r=100\) for our method. The overall performance of our method compared to baseline methods is summarized in Table~\ref{tab:qr}. Our method achieves higher \textbf{NDCG@\{10,20\}} and \textbf{Recall@\{10,20\}} scores, along with smaller parameter sizes across all three datasets, confirming the effectiveness of our approach in refining coarse meta-embeddings and enhancing compositional full embeddings (see Sec.\ref{sec:refine}).

In contrast, variable-size methods such as \textbf{PEP}, \textbf{QR}, and \textbf{LEGCF} perform suboptimally as they overlook fine-grained semantics, resulting in less effective compositional full embeddings for user/item representation. Among the one-size methods, \textbf{NimbleTT} and \textbf{CIESS} have similar parameter sizes to the variable-size methods but do not match our performance. Techniques like \textbf{OptEmbed}, \textbf{ESAPN}, and \textbf{AutoEmb} produce embedding layers with significantly larger parameter sizes compared to ours. \textbf{DHE} shows the poorest performance on all datasets, indicating that small dimension size hash codes fail to enhance embedding uniqueness.

Compared to the unified dimensionality (\textbf{UD}) settings, our method demonstrates a superior balance between performance and parameter efficiency across the Gowalla, Yelp2020, and Amazon-book datasets. Our method consistently achieves competitive performance while significantly reducing parameter size. This efficiency confirms that our approach effectively refines coarse meta-embeddings into high-quality compositional full embeddings, making it suitable for scenarios with strict memory constraints. The substantial reduction in parameter size without a proportional drop in performance highlights the robustness and practicality of our method in diverse applications.

To evaluate the statistical significance of performance improvements between the baseline model (LEGCF) and our proposed model, we employ p-values. A p-value quantifies the probability of observing differences as extreme as those found, assuming no true difference exists between the models. Typically, a threshold of \textbf{0.05} is used: if the p-value falls below \textbf{0.05}, we can reject the null hypothesis, concluding that the difference is statistically significant. If the p-value is between \textbf{0.05} and \textbf{0.1}, it indicates marginal significance, suggesting a trend that merits further investigation. Conversely, a p-value exceeding \textbf{0.1} suggests insufficient evidence to claim a statistically significant difference, indicating that observed results may be due to random chance. To calculate the p-value, we conduct multiple experiments and compare performance metrics (such as NDCG and Recall) between the two models. We utilize statistical tests, such as paired t-tests, to compute the p-value based on observed performance differences across various runs or seeds. As shown in Table.\ref{p_value}, we can observe that for the Gowalla and Yelp2020 datasets, with all significant p-values below \textbf{0.05} for most comparisons, confirming statistical significance. Specifically, the improvements in performance for the Gowalla dataset are highly significant, while the results for the Yelp2020 dataset also show promising trends, though one result is marginally significant.

\begin{figure*}
  \centering
  \includegraphics[width=\textwidth]{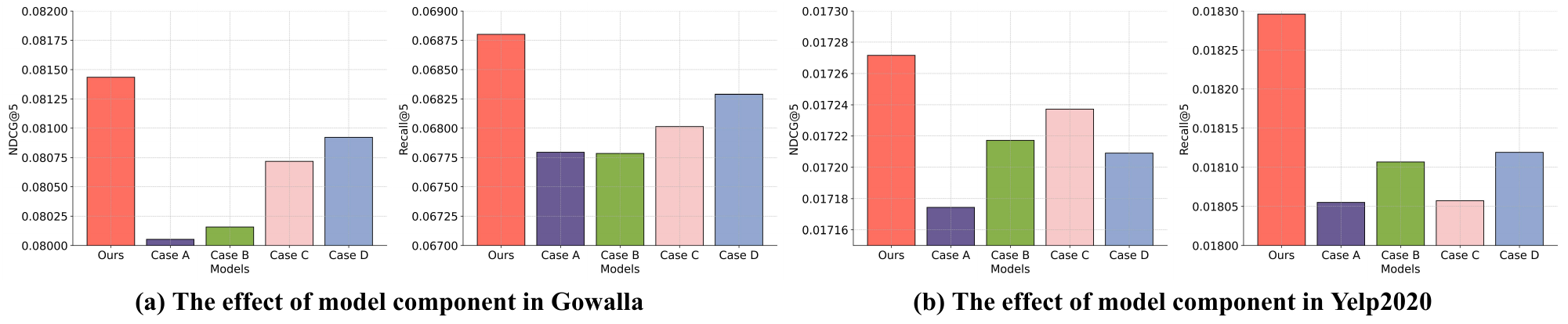}
  \caption{The performance of our method on w.r.t. various model components.}\label{fig:model-ab}
\end{figure*}

\subsection{Model Component Analysis (Q2)}
To validate the performance gain from each key component, we conduct the ablation studies on the several innovative components including the fine meta-embedding initialization strategy, soft thresholding sparsity strategy and weight bridging updating strategy. The performance comparison is depicted in Fig.\ref{fig:model-ab}

\noindent (1) \textbf{Fine Meta-Embedding Initialization}: We enhance the initialization strategy for the fine meta-embedding \(\textbf{E}^r_{meta}\) in two ways. First, instead of randomly initializing the embedding values, we select a portion of the coarse meta-embedding to form \(\widetilde{\textbf{E}}^c_{meta}\). Second, we use SparsePCA to preserve the main patterns and control sparsity. Our experiments show that the absence of these two components (\textbf{Case A and B}) leads to significantly worse performance. This indicates that initializing \(\textbf{E}^r_{meta}\) with semantic and sparse properties is crucial for training an accurate meta-embedding codebook, resulting in high-quality entity embeddings.  Such fact further discloses our core contribution -- \textit{refining the coarse meta-embedding via the fine-gained semantic information} (Sec.\ref{sec:refine})

\noindent (2) \textbf{Soft Thresholding Sparsity Strategy}: To evaluate the effect of soft thresholding, we removed that and directly used \(\textbf{E}^r_{meta}\) to refine \(\textbf{E}^c_{meta}\) (\textbf{Case C}). The resulting performance decline without soft thresholding demonstrates that \textit{soft thresholding technique is integrated to dynamically adjust the sparsity levels by smoothly driving small coefficients to zero while retaining the significant ones. This method ensures that the resulting embeddings remain sparse and computationally efficient, facilitating better convergence in training processes}, as claimed in Sec.\ref{sec:refine}.

\noindent (3) \textbf{Weight Bridging Updating Strategy}: We leverage the Eq.\ref{eq:liaw} to calculate \(\textbf{S}^r\), we abandon that and match these two kinds of meta-embedding without considering the users/items' full embedding to validate the importance of our bridging strategy (\textbf{Case D}). Our proposed strategy aims at well \textit{matching each coarse meta-embedding with several fine meta-embeddings according to users/items' semantics } (Sec.\ref{sec:liaw}). Fig.\ref{fig:model-ab} proves that.

\subsection{Hyperparameter Analysis (Q3)} \label{sec:ha}
To study the sensitiveness of our work on hyperparameters, we conduct hyperparameter analysis on the desired number of new features $d^r$ in SparsePCA, the threshold parameter $\lambda$ in soft thresholding, the weight of the fine meta-embedding $w_{cr}$ and the number of fine meta-embedding assigned to each coarse meta-embedding $t^r$.

\begin{figure*}
  \centering
  \includegraphics[width=\textwidth]{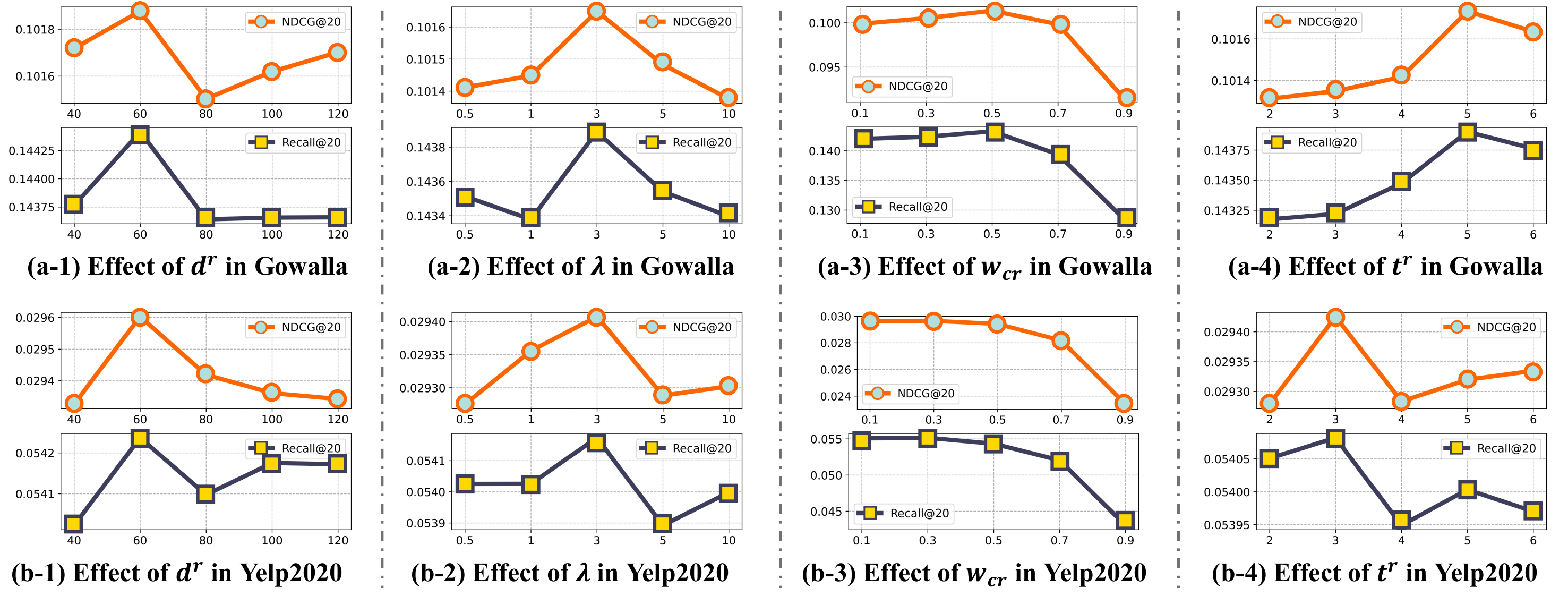}
  \caption{The performance of our method on w.r.t. various hyperparameter settings.}\label{fig:ab}
\end{figure*}

\noindent (1) \textbf{Number of New Features $d^r$}: The embedding size \( d \) for both \( \textbf{E}^c_{meta} \) and \( \textbf{E}^r_{meta} \) is set to 128. For testing, we set the \( d^r \) values to \{40; 60; 80; 100; 120\}. The performance results on both datasets are shown in Fig. \ref{fig:ab}(a-1) and (b-1). It was found that increasing the number of non-zero values does not guarantee performance improvement. This is primarily because the fine stage focuses on learning fine-grained semantic information. A densely fine meta-embedding may capture redundant semantics similar to those in the coarse stage, thereby neglecting finer details. Conversely, embeddings that are too sparse (e.g., \( d^r = 40 \)) can impair the ability to express semantic nuances effectively. By experimenting with different levels of sparsity for initialization, we aim to balance capturing detailed semantics and avoiding redundancy.

\noindent (2) \textbf{Threshold of Soft Thresholding \(\lambda\)}: To investigate the impact of the soft thresholding parameter \(\lambda\) on embedding sparsity and representation quality, we conducted experiments with \(\lambda\) values set to \{0.5; 1; 3; 5; 10\}. These values determine different ratios of non-zero values in the fine meta-embedding. As shown in Fig. \ref{fig:ab}(a-2) and (b-2), our method achieves optimal performance on both datasets when the fine meta-embedding remains sparse. Specifically, for Gowalla and Yelp2020, the best performance is observed with \(\lambda = 3\), which results in approximately half of the values being non-zero. This outcome aligns with our previous findings regarding the hyperparameter \(d^r\), further validating that maintaining an optimal level of sparsity is crucial for capturing fine-grained semantic information without redundancy. Our results can also indicate that the refined stage is designed to work under tight memory budget.

\noindent (3) \textbf{Weight of the Fine Meta-Embedding $w_{cr}$}:The chart showing the balance between the coarse meta-embedding and the fine meta-embedding is depicted in Fig.\ref{fig:ab}(a-3) and (b-3). The set of \( w_{cr} \) values tested includes \{0.1, 0.3, 0.5, 0.7, 0.9\}. It is observed that for the Gowalla dataset, the best result is achieved with a weight of 0.5, while for Yelp2020, the optimal \( w_{cr} \) is 0.3. As the weight decreases, the performance becomes less competitive. This phenomenon indicates that the fine stage is essential for enhancing the coarse meta-embedding, and that the sparse fine meta-embedding performs well even under tight memory constraints.

\noindent (4) \textbf{Number of Fine Meta-Embedding $t^r$ per Coarse Meta-Embedding}: The performance regarding the number of fine meta-embeddings assigned to each coarse meta-embedding is plotted in Fig. \ref{fig:ab}(a-4) and (b-4). The values for \( t^r \) are from the set \{2, 3, 4, 5, 6\}. It is observed that for the Gowalla dataset, our method achieves the best performance with the larger setting \( t^r = 5 \), while for Yelp2020, the best performance is achieved with \( t^r = 3 \). This phenomenon can be attributed to the fact that the larger dataset adopts a deeper GNN for training, allowing the fine meta-embeddings to have stronger representational capability. As a result, fewer fine meta-embeddings are sufficient to effectively refine the coarse embeddings. Therefore, smaller datasets may require more fine meta-embeddings to achieve optimal performance.

\subsection{Space Complexity (Q4)}
\begin{figure*}
  \centering
  \includegraphics[width=\textwidth]{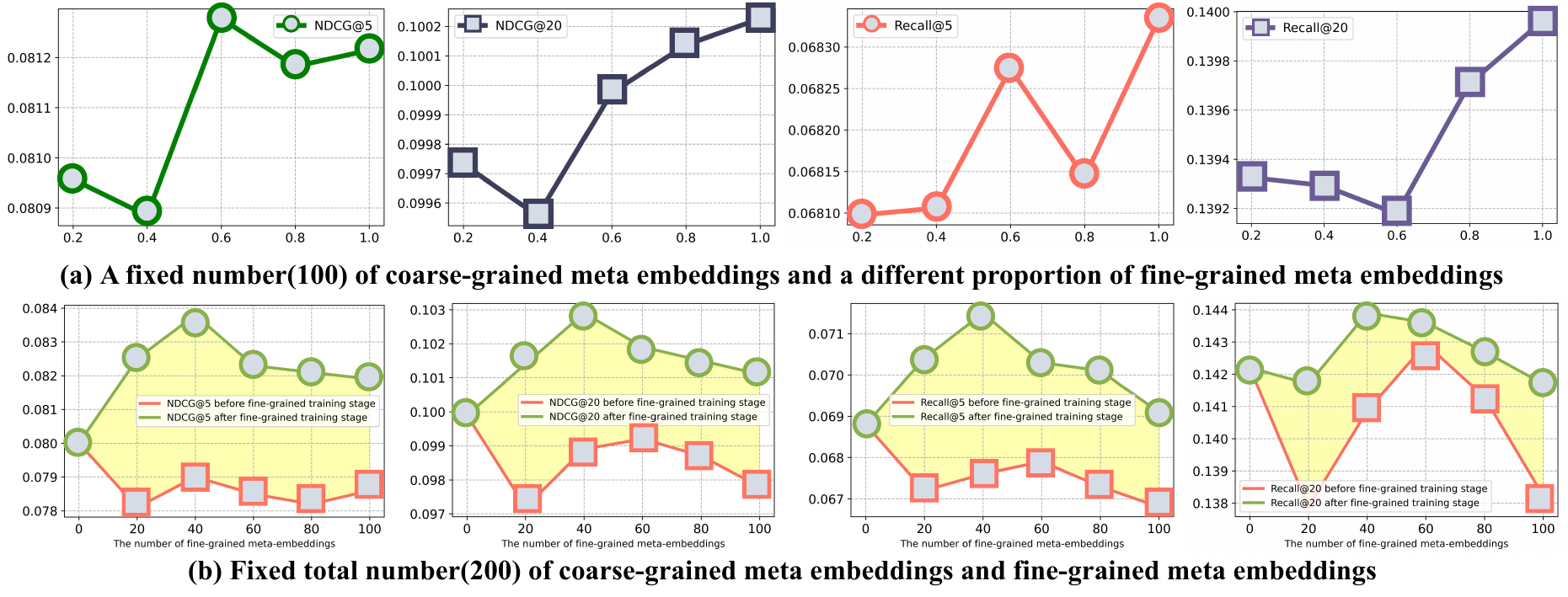}
  \caption{The performance of our method on w.r.t. various ratios of the coarse meta-embeddings and the fine meta-embeddings.}\label{fig:ratio}
\end{figure*}

We evaluate the memory consumption of our framework by adjusting the ratios between the coarse meta-embeddings and the fine ones. In LEGCF, the space complexity is $O(tN + md)$ for its lightweight embeddings, where $t$ is the number of meta-embeddings per entity and $m$ is the number of meta-embedding codebooks. Compared to the space complexity of a full embedding table, where $t \ll m$, we have $O(tN + md) \ll O(Nd)$. In our method, the space complexity of the coarse stage is $O(t^cN + m^cd)$, similar to LEGCF. However, in the fine stage, as discussed in Section \ref{sec:ha}, setting half of the fine meta-embedding values to zero achieves the best performance. Therefore, the space complexity for this stage can be calculated as $O(t^rm^c + \frac{m^rd}{2})$. The overall complexity is $O(t^cN + t^rm^c + \frac{m^r + 2m^c}{2}d) < O(tN + md)$, given that $m^c + m^r < m$, $m^r \leq m^c$, and $t^r \ll d$ in our setting. Under a fixed memory constraint, a lower $m^c$ results in lower space complexity. The fine stage helps to balance the trade-off between embedding uniqueness and embedding fidelity. Fig.\ref{fig:ratio} further demonstrates that.

We conducted two types of ablation studies to explore how to set the number of coarse-grained and fine-grained meta-embeddings to reduce space complexity while achieving better performance. As shown in Fig.\ref{fig:ratio}(a), based on a fixed number of coarse-grained meta-embeddings, we adjusted the ratios of fine-grained meta-embeddings. The results show that because the coarse-grained training stage effectively establishes strong correlations between entities and coarse-grained meta-embeddings, increasing the number of fine-grained meta-embeddings significantly improves performance. However, when space capacity is strictly limited, fewer coarse-grained meta-embeddings cannot adequately balance embedding uniqueness and fidelity. In such cases, a smaller ratio of fine-grained meta-embeddings may perform better. As illustrated in Fig.\ref{fig:ratio}(b), we conducted experiments with a fixed total number of coarse-grained and fine-grained meta-embeddings to validate this observation.

\section{Conclusion}
In this paper, we propose a novel GNN-based model for multi-grained semantic learning of lightweight meta-embeddings for ID-based recommendation systems. This hierarchical and selective connectivity ensures our model to learn wide-ranging semantic information at the coarse level, while focusing on granular, personalized features at the fine level. Our contributions include a two-tiered virtual node structure that captures coarse-grained and fine-grained semantics, an innovative initialization approach relying on SparsePCA comes up to preserve the sparsity and maintain associative relationships, dynamic soft thresholding to adjust sparsity levels during training process, and a weight bridging update strategy that aligns coarse-grained and fine-grained meta-embeddings based on semantic relevance.  Experimental results over benchmarks validate the advantages of our model over state-of-the-art counterparts.

\Acknowledgements{This work was supported in part by the Research Projects of the National Natural Science Foundation of China under Grant U21A20470, Grant 62172136, and Grant 72188101 and in part by the Institute of Advanced Medicine and Frontier Technology under Grant 2023 IHM01080.}






\begin{thebibliography}{99}
\bibitem{2020Chen} Chen T, Yin H, Ye G, et al. Try this instead: Personalized and interpretable substitute recommendation. In: Proceedings of the 43rd international ACM SIGIR conference on research and development in information retrieval, 2020. 891-900.
\bibitem{2022Chen} Chen T, Yin H, Long J, et al. Thinking inside the box: learning hypercube representations for group recommendation. In: Proceedings of the 45th International ACM SIGIR Conference on Research and Development in Information Retrieval, 2022. 1664-1673.
\bibitem{2021Zhang} Zhang C, Wang Y, Zhu L, et al. Multi-graph heterogeneous interaction fusion for social recommendation. ACM Transactions on Information Systems, 2021, 40(2): 1-26
\bibitem{2022Wang} Wang Y, Peng J J, Wang H B, et al. Progressive learning with multi-scale attention network for cross-domain vehicle re-identification. Sci China Inf Sci, 2022, 65(6): 160103, https://doi.org/10.1007/s11432-021-3383-y
\bibitem{2021multi-modal} Wang Y. Survey on deep multi-modal data analytics: Collaboration, rivalry, and fusion. ACM Transactions on Multimedia Computing, Communications, and Applications (TOMM), 2021, 17(1s): 1-25
\bibitem{2022Bayesian} Chen Y, Wang Y, Ren P, et al. Bayesian feature interaction selection for factorization machines. Artificial Intelligence, 2022, 302: 103589
\bibitem{FM} Rendle S. Factorization machines. In: Proceedings of the IEEE International Conference on Data Mining, 2010. 995-1000
\bibitem{NCF} He X, Liao L, Zhang H, et al. Neural Collaborative Filtering. In: Proceedings of the 26th International Conference on World Wide Web, 2017. 173-182
\bibitem{Lightgcn} He X, Deng K, Wang X, et al. Lightgcn: Simplifying and powering graph convolution network for recommendation. In: Proceedings of the 43rd International ACM SIGIR Conference on Research and Development in Information Retrieval, 2020. 639-648
\bibitem{PEP} Liu S, Gao C, Chen Y, et al. Learnable embedding sizes for recommender systems. 2021. ArXiv:2101.07577
\bibitem{AutoEmb} Zhaok X, Liu H, Fan W, et al. Autoemb: Automated embedding dimensionality search in streaming recommendations. In: Proceedings of the 21st IEEE International Conference on Data Mining, 2021. 896-905
\bibitem{ESAPN} Liu H, Zhao X, Wang C, et al. Automated embedding size search in deep recommender systems. In: Proceedings of the 43rd International ACM SIGIR Conference on Research and Development in Information Retrieval, 2020. 2307-2316
\bibitem{OptEmbed} Lyu F, Tang X, Zhu H, et al. Optembed: Learning optimal embedding table for click-through rate prediction. In: Proceedings of the 31st ACM International Conference on Information \& Knowledge Management, 2022. 1399-1409
\bibitem{CIESS} Qu Y, Chen T, Zhao X, et al. Continuous input embedding size search for recommender systems. In: Proceedings of the 46th International ACM SIGIR Conference on Research and Development in Information Retrieval, 2023. 708-717
\bibitem{DHE} Kang W C, Cheng D Z, Yao T, et al. Learning to embed categorical features without embedding tables for recommendation. In: Proceedings of the 27th ACM SIGKDD Conference on Knowledge Discovery \& Data Mining, 2021. 840-850
\bibitem{NimbleTT} Yin C, Zheng D, Nisa I, et al. Nimble gnn embedding with tensor-train decomposition. In: Proceedings of the 28th ACM SIGKDD Conference on Knowledge Discovery and Data Mining, 2022. 2327-2335
\bibitem{EODR} Xia X, Yu J, Wang Q, et al. Efficient on-device session-based recommendation. ACM Transactions on Information Systems, 2023, 41(4): 1-24
\bibitem{PPSR} Han J, Ma Y, Mei Q, et al. Deeprec: On-device deep learning for privacy-preserving sequential recommendation in mobile commerce. In: Proceedings of the Web Conference, 2021. 900-911
\bibitem{NPOIR} Wang Q, Yin H, Chen T, et al. Next point-of-interest recommendation on resource-constrained mobile devices. In: Proceedings of the Web conference, 2020. 906-916
\bibitem{2024Long} Long J, Ye G, Chen T, et al. Diffusion-Based Cloud-Edge-Device Collaborative Learning for Next POI Recommendations. 2024. ArXiv:2405.13811
\bibitem{LEE} Chen T, Yin H, Zheng Y, et al. Learning elastic embeddings for customizing on-device recommenders. In: Proceedings of the 27th ACM SIGKDD Conference on Knowledge Discovery \& Data Mining, 2021. 138-147
 \bibitem{PEEL} Zheng R, Qu L, Chen T, et al. Personalized Elastic Embedding Learning for On-Device Recommendation. IEEE Transactions on Knowledge and Data Engineering, 2024.
\bibitem{ODNIR} Xia X, Yin H, Yu J, et al. On-device next-item recommendation with self-supervised knowledge distillation. In: Proceedings of the 45th International ACM SIGIR Conference on Research and Development in Information Retrieval, 2022. 546-555
\bibitem{2023Qian} Qian B, Wang Y, Hong R, et al. Adaptive data-free quantization. In: Proceedings of the IEEE/CVF Conference on Computer Vision and Pattern Recognition, 2023. 7960-7968
\bibitem{2022Qian} Qian B, Wang Y, Yin H, et al. Switchable online knowledge distillation. In: Proceedings of the European Conference on Computer Vision, 2022. 449-466
\bibitem{2024Unpacking} Wang Y, Qian B, Liu H, et al. Unpacking the gap box against data-free knowledge distillation. IEEE Transactions on Pattern Analysis and Machine Intelligence, 2024.
    \bibitem{QR} Shi H J M, Mudigere D, Naumov M, et al. Compositional embeddings using complementary partitions for memory-efficient recommendation systems. In: Proceedings of the 26th ACM SIGKDD International Conference on Knowledge Discovery \& Data Mining, 2020. 165-175
\bibitem{FBDH} Zhang C, Liu Y, Xie Y, et al. Model size reduction using frequency based double hashing for recommender systems. In: Proceedings of the 14th ACM Conference on Recommender Systems, 2020. 521-526
\bibitem{CEL} Chen Y, Huzhang G, Zeng A, et al. Clustered Embedding Learning for Recommender Systems. In: Proceedings of the Web Conference, 2023. 1074-1084
\bibitem{LCCE} Liang X, Chen T, Nguyen Q V H, et al. Learning compact compositional embeddings via regularized pruning for recommendation. In: Proceedings of the IEEE International Conference on Data Mining, 2023. 378-387
\bibitem{LightweightEmbeddings} Liang X, Chen T, Cui L, et al. Lightweight Embeddings for Graph Collaborative Filtering. In: Proceeding of the 47th International ACM SIGIR Conference on Research and Development in Information Retrieval. 2024
\bibitem{BPR} Rendle S, Freudenthaler C, Gantner Z, et al. BPR: Bayesian personalized ranking from implicit feedback. 2012. ArXiv:1205.2618
\bibitem{GeneralizedInverse} Meyer, Jr C D. Generalized Inverses (Theory And Applications) (Adi Ben-Israel and Thomas NE Greville). SIAM Review, 1976, 18(2): 320-322
\bibitem{Neuralgraph} Wang X, He X, Wang M, et al. Neural graph collaborative filtering. In: Proceedings of the 42nd International ACM SIGIR Conference on Research and Development in Information Retrieval, 2019. 165-174
\bibitem{gowalla} Cho E, Myers S A, Leskovec J. Friendship and mobility: user movement in location-based social networks. In: Proceedings of the 17th ACM SIGKDD international conference on Knowledge discovery and data mining, 2011. 1082-1090
\bibitem{amazon-book} He R, McAuley J. Ups and downs: Modeling the visual evolution of fashion trends with one-class collaborative filtering. In: proceedings of the 25th international conference on world wide web, 2016. 507-517
\bibitem{METIS} Karypis G, Kumar V. METIS: A software package for partitioning unstructured graphs, partitioning meshes, and computing fill-reducing orderings of sparse matrices. 1997.
\bibitem{DGTS} Liu H, Wang Y, Wang M, et al. Delving globally into texture and structure for image inpainting. In: Proceedings of the 30th ACM International Conference on Multimedia. 2022: 1270-1278.
\bibitem{UnderstandDNN} Glorot X, Bengio Y. Understanding the difficulty of training deep feedforward neural networks. In: Proceedings of the thirteenth international conference on artificial intelligence and statistics, 2010. 249-256
\bibitem{Adam} Kingma D P. Adam: A method for stochastic optimization. 2014. ArXiv:1412.6980.
\bibitem{Yelp} Alam M, Cevallos B, Flores O, et al. Yelp Dataset Analysis using Scalable Big Data. 2021. ArXiv:2104.08396.
\bibitem{SPCA} Zou H, Hastie T, Tibshirani R. Sparse principal component analysis. J Computational and Graphical Statistics, 2006, 15(2): 265-286
\bibitem{On-deviceRS} Yin H, Qu L, Chen T, et al. On-device recommender systems: A comprehensive survey. 2024. ArXiv:2401.11441.
\bibitem{Semi} Qu L, Tang N, Zheng R, et al. Semi-decentralized federated ego graph learning for recommendation. In: Proceedings of the Web Conference, 2023. 339-348.

\end{thebibliography}
\end{document}